\documentclass[journal]{IEEEtran}
\usepackage{amsmath,amsfonts}
\usepackage{algorithmic}
\usepackage{array}
\usepackage{textcomp}
\usepackage{stfloats}
\usepackage{url}
\usepackage{verbatim}
\usepackage{graphicx}
\def\BibTeX{{\rm B\kern-.05em{\sc i\kern-.025em b}\kern-.08em
    T\kern-.1667em\lower.7ex\hbox{E}\kern-.125emX}}
\usepackage{balance}

\usepackage{makecell} 
\usepackage{soul}  
\usepackage{multirow}
\usepackage[table]{xcolor}
\usepackage{pifont}

\usepackage{array,booktabs}   
\usepackage{graphicx} 
\usepackage{subcaption}

\definecolor{tab-green}{RGB}{34,139,34} 
\definecolor{tab-orange}{RGB}{204,85,0} 
\definecolor{tab-blue}{RGB}{0,0,255}

\usepackage{tcolorbox}
\usepackage{hyperref}
\usepackage[numbers,sort&compress]{natbib}

\begin{document}

\title{Are They All Good? Evaluating the Quality of CoTs in LLM-based Code Generation}

\author{Binquan Zhang, Li Zhang, Zhiwen Luo, Yuxin Du, Fang Liu, Song Wang, Lin Shi

\thanks{B. Zhang, L. Zhang, Z. Luo, Y. Du, F. Liu, and L. Shi are with Beihang University, China. Emails are: {binquan,lily,zhiwenluo,duyuxn,fangliu,shilin}@buaa.edu.cn

S. Wang is with York University, Canada. Email is: wangsong@yorku.ca
}}

\markboth{Journal of \LaTeX\ Class Files,~Vol.~18, No.~9, September~2020}%
{How to Use the IEEEtran \LaTeX \ Templates}

\maketitle

\begin{abstract}

Large language models (LLMs) have demonstrated impressive performance in code generation, particularly when augmented with chain-of-thought (CoT) prompting techniques. 
They break down requirements into intermediate reasoning steps, which act as design rationales to guide LLMs in writing code like human programmers. 
Thus, the quality of these steps is crucial for ensuring the correctness and reliability of the generated code.
However, little is known about the quality of CoT generated by LLMs.
To what extent can we trust the thoughts generated by LLMs? How good are they? 
This paper empirically explores the external and internal factors of why LLMs generate unsatisfactory CoTs by analyzing 1,023 failed code samples on two widely used code generation benchmarks.
We also evaluate their impact on code generation performance by analyzing 210 CoT-code pairs and refining the unsatisfied CoTs by prompting LLMs.
Our study yields the following findings: 
1)  Among the factors affecting CoT quality, external factors account for 53.60\%, primarily including unclear requirements and lack of contextual information.
Internal factors make up 40.10\%, mainly due to inconsistencies between CoT and prompts caused by LLMs' misunderstanding of the instructions. 
2) Despite CoT being correct, 18.5\% of the generated code still contains errors.
This is primarily due to LLMs failing to follow instructions, leading to inconsistencies between CoT and the code.
Additionally, we found that even when the code is correct, there is an 11.90\% chance that the CoT contains errors.
3) Our further research on refining the low-quality CoTs reveals that LLMs can improve CoT, especially when providing detailed CoT problem information.
Our findings shed light on the underlying issues that hinder the effectiveness of CoT in LLM-based code generation, offering valuable insights for enhancing both the reasoning process and the overall reliability of code generation.


\end{abstract}


\section{Introduction}







Large Language Models (LLMs) have recently made remarkable advancements in their reasoning capabilities, a trend exemplified by the emergence of reasoning models like OpenAI's o1~\cite{openai}, DeepSeek-R1~\cite{guo2025deepseek}, and Gemini-2.0-Flash-Thinking~\cite{gemini}.
Chain of Thought (CoT)~\cite{wei2022chain} represents one of the most pivotal breakthroughs in eliciting these reasoning abilities. 
This approach guides models to systematically articulate their thought processes before delivering a final answer, effectively bridging the gap between ``fast thinking'' and ``slow thinking''. 
In complex domains like code generation, this step-by-step reasoning is particularly valuable.
It emulates the cognitive strategies of human programmers by ``forcing'' LLMs to break down complex task requirements into a series of intermediate reasoning steps~\cite{li2023structured,ma2023bridging,huang2024codecottacklingcodesyntax}. 
These steps serve as a design foundation, guiding LLMs to write code like human programmers.
Extensive research has been done to generate multi-step reasoning chains with LLMs, such as Tree-of-Thought(ToT)~\cite{NEURIPS2023_271db992}, Structured Chain-of-Thought(SCoTs)~\cite{li2023structured}, Reasoning-via-Planning(RAP)~\cite{hao2023reasoning}, among others~\cite{xie2023self,zhuang2023toolchain,khalifa2023grace}.
While these methods have improved the performance of code generation, little is known about the quality of the intermediate reasoning steps.
\textbf{\textit{To what extent can we trust the ``thoughts'' generated by LLMs? How good are they?} }

Therefore, we conduct a large-scale empirical study that evaluates the quality of CoTs and explores automated methods for their detection and repair.
We first collect CoTs generated by LLMs and the corresponding code produced based on these CoTs from two code generation benchmarks(i.e., CoderEval~\cite {Zhang_2024} and SWE-bench~\cite{jimenez2023swe}), resulting in a total of 1,023 CoT-code pairs.
Among these pairs, we then identify 813 code samples that fail to pass the tests and conduct an in-depth analysis of the associated CoTs.
Since CoTs are generated by LLMs in response to a given requirement, we categorize the root causes of low-quality CoTs into two main aspects: external and internal factors, where external factors refer to issues arising from the input requirement during CoT generation, while internal factors are relate to the content of the CoT itself. 
Our empirical study identifies external factors such as \textbf{unclear implementation details} and \textbf{missing contextual information} in the prompt. Internal factors include \textbf{misunderstanding explicit requirements}, \textbf{incomprehension of implicit requirements}, and \textbf{incorrect planning} within the CoT.  
Upon evaluation, we find that out of the 813 LLM-generated CoTs, 53.6\% of the quality issues stem from external factors, while 40.1\% are attributed to internal factors.


Beyond analyzing the 813 CoTs, we also extend our investigation to the remaining 210 passing code samples and their associated CoTs.
This additional analysis provides a more comprehensive understanding of the impact of LLM-generated CoTs on code quality. 
Our findings reveal that \textit{even when the CoT is correct, 18.5\% of the generated code still contains errors}.
This issue arises mainly because LLMs sometimes fail to adhere to the steps in the CoT, resulting in discrepancies between the CoT and the code.
Furthermore, our analysis reveals that \textit{even when the generated code is correct, there is still an 11.9\% chance of an incorrect CoT}.
This occurs because LLMs possess strong adaptive capabilities.
Even if certain steps in the CoT are flawed, LLMs can rely on its inherent knowledge and contextual understanding to correct these errors and ultimately generate accurate code.

Finally, we investigate the self-detection and self-repair capabilities of LLMs. 
For the detection of faulty CoTs, we employ a Multi-Agent Debate (MAD)~\cite{liang2023encouraging} framework. 
The experiments demonstrate that the MAD framework significantly outperforms a single LLM.
Regarding the repair of faulty CoTs, we examine the impact of feedback information at different granularities (ranging from simple feedback to error type information and detailed error descriptions) on the self-repair capabilities of LLMs. 
Our findings reveal a clear positive correlation: the more detailed the provided feedback, the more effective the model's repair performance. While this is a preliminary study, we believe that LLMs hold significant potential in detecting and repairing erroneous CoTs.

To summarize, this paper makes the following contributions:
\begin{itemize}

\item To the best of our knowledge, we conduct the first in-depth examination of the factors influencing the quality of LLM-generated CoTs and propose a comprehensive taxonomy of CoT errors, encompassing both external and internal factors.

\item We further investigate the relationship between CoT quality and the correctness of the generated code. Our findings indicate that even correct CoTs can sometimes lead to faulty code, while, conversely, a small portion of correct code can still be generated from incorrect CoTs.

\item {We conduct a preliminary exploration of faulty CoT detection and repair strategies across different LLMs, highlighting that more detailed error information improves LLMs' performance in refining faulty CoTs.}

\item Our findings shed light on the underlying issues that hinder the effectiveness of CoT in LLM-based code generation, offering valuable insights for enhancing both the reasoning process and the overall reliability of generated code.

\end{itemize}
\section{Background \& Related Work}

\subsection{Code Generation with LLMs}
\label{code generation}
LLMs have recently gained prominence in Software Engineering (SE) for automatic code generation.
Among the most notable is OpenAI’s CodeX~\cite{chen2021evaluating}, a GPT-based auto-regressive model with up to 12 billion parameters, fine-tuned on 54 million public GitHub repositories.
CodeX powers \href{https://github.com/features/copilot}{GitHub Copilot}~\cite{asare2023github}, an IDE assistant capable of generating code based on user-provided context, supporting up to 4,096 tokens of input.
A key enhancement to Copilot is ``\href{https://docs.github.com/en/copilot/responsible-use-of-github-copilot-features/responsible-use-of-github-copilot-chat-in-your-ide}{Copilot Chat}'', which integrates human feedback to excel at a range of coding tasks, from code generation~\cite{vaithilingam2022expectation} to bug fixing~\cite{asare2023github} and explaining code in natural language~\cite{wermelinger2023using}.

In addition to CodeX, Meta’s Code Llama~\cite{2024codellamaopenfoundation} has emerged as another leading model for code generation.
Built on the Llama 2~\cite{touvron2023llama2openfoundation} framework, Code Llama supports various programming languages and excels in tasks like bug fixing, code explanation, and code generation with models ranging from 7 to 34 billion parameters.
Other notable models in this space include Salesforce’s CodeRL~\cite{le2022coderlmasteringcodegeneration} and CodeGen~\cite{nijkamp2023codegenopenlargelanguage}, and Amazon’s CodeWhisperer~\cite{codewhisperer}, all contributing to the growing ecosystem of LLM-based code generation tools.

Recent advancements in code generation often utilize the CoT framework, where LLMs first generate reasoning chains before producing the final code. 
Li et al.~\cite{li2023structured} introduced Structured Chain-of-Thought Prompting (SCoT), which incorporates programming structures such as sequences and loops to reduce errors and improve the understanding of requirements.
Jiang et al.~\cite{jiang2024selfplanningcodegenerationlarge} developed a planning-based approach that separates the planning phase, where CoTs are created, from the implementation phase, where code is generated step-by-step.
Huang et al.'s CodeCoT~\cite{huang2024codecottacklingcodesyntax} framework improves code correctness through self-checking and test case generation, while Le et al.'s~\cite{le2024codechainmodularcodegeneration} CodeChain method enhances modularity and maintainability by employing submodule reuse and self-revision.

While LLM-based code generation techniques have improved the accuracy and reliability of generated code, the extent of these improvements remains limited.
The core issue stems from the critical influence that the quality of the CoT has on the final output.
Any missteps in the CoT can lead to code that fails to meet the original requirements or introduces errors.
However, there has been insufficient focus on effectively evaluating and refining the quality of LLM-generated CoT, which constrains the performance of current methods in code generation tasks. 
Thus, the rigorous evaluation and refinement of CoT quality is a crucial challenge for advancing code generation techniques further.
Given these challenges, we systematically assessed LLM-generated CoT quality on two widely used code generation datasets and explored their potential to improve flawed CoTs.

\subsection{Chain of Thought}
For reasoning tasks, directly requesting an answer from an LLM is often inadequate~\cite{yugeswardeenoo2024question,qi2024mutual,abbasiantaeb2024let}.
Kojima et al.~\cite{kojima2022large} discovered that appending the phrase ``let's think step by step'' to the end of the prompt enables GPT-3~\cite{brown2020languagemodelsfewshotlearners} to perform multi-step reasoning in a zero-shot setting, ultimately leading to the correct answer.
Wei et al.~\cite{wei2022chain} later introduced CoT prompting, which involves using an LLM to generate a series of intermediate reasoning steps to improve performance in complex reasoning tasks.

Since code generation often involves multiple logical steps and complex reasoning processes, the concept of CoT prompting is particularly well-suited for the code generation task~\cite{suzgun2022challenging}.
By using LLM-generated CoT, an LLM can approach problem-solving like a human programmer~\cite{yang2024swe, xia2024agentless}, incrementally building solutions step by step, which leads to improvements in the correctness and reliability of the generated code.
Many studies~\cite{li2023structured,huang2024codecottacklingcodesyntax} have demonstrated the effectiveness of LLM-generated CoT in improving code generation performance.
For example, \citet{li2023structured} showed that LLM-generated structured CoTs help enhance generated code quality.
\citet{huang2024codecottacklingcodesyntax} introduced a CoT framework with a self-examination component to improve the robustness of the generated code, particularly in handling syntax errors.
However, there is a dearth of research concerning the quality of LLM-generated CoTs in the context of code generation.
LLM-generated CoT with quality issues can hinder improvements in code generation performance, potentially compromising the correctness and reliability of the generated code.

\subsection{Evaluation of LLM-Generated Code Quality}
As LLMs are increasingly used, even by users without programming backgrounds, the quality of generated code has become a growing concern, prompting researchers to address potential risks~\cite{spiess2024quality,gu2024testart}.
Thus, systematically studying LLM-generated code quality and exploring solutions has become a critical research direction.

\citet{liu2024refining} conducted the first systematic evaluation of the reliability of code generated by ChatGPT, filling a significant research gap in this area. They created a dataset comprising 2,033 programming tasks and 4,066 code snippets generated by ChatGPT. Their analysis not only assessed the correctness of the code but also explored other quality attributes, such as maintainability, providing new insights into the challenges associated with ChatGPT-generated code. \citet{tambon2024bugslargelanguagemodels} utilized the CoderEval dataset to identify and classify defective Python code snippets generated by three different LLMs. By systematically analyzing error patterns, they proposed a detailed error classification framework that offers a valuable reference for future research and practice. Additionally, they gathered feedback and handling experiences from developers through surveys. \citet{liu2024needliftfingeranymore}conducted a comprehensive evaluation of ChatGPT's performance on programming problems of varying difficulty levels, including accuracy, runtime, and memory overhead. Their study revealed the limitations of ChatGPT when handling novel problems, offering new insights into its capability boundaries. Kevin \citet{10174227} investigated the application of LLMs in programming, particularly focusing on how to avoid simple programming errors. By comparing the impact of different prompting techniques on Codex performance, they provided a new perspective on understanding LLM behavior. \citet{liu2023codegeneratedchatgptreally} proposed the ``EvalPlus'' evaluation framework, which combines LLMs with mutation strategies to generate a large number of test cases. This approach enhanced existing benchmark datasets (such as HUMANEVAL) and created HUMANEVAL+, significantly improving the assessment of functional correctness in LLM-generated code. \citet{huang2024codecottacklingcodesyntax} introduced the ``Self-exam CodeCoT'' framework, which improves code reliability through self-checking and iterative refinement, allowing LLM-generated code to automatically detect and correct errors. This framework also enhances user trust in automatically generated code through a transparent testing and revision process.

These studies highlight new methods and perspectives for addressing quality and reliability issues in LLM-generated code, revealing the challenges faced by single models in covering all programming scenarios and providing valuable insights for further optimization of LLMs.


\section{Taxonomy of Factors Influencing CoT Quality}

This section outlines the experimental design, taxonomy of CoT quality factors, and analyzes external and internal factors.

\subsection{Study Design}
\label{sec: study design}

\subsubsection{The studied LLMs}
We selected three cutting-edge models renowned for their advanced reasoning and code generation capabilities as our research subjects:
Gemini-2.0-Flash-Thinking-Exp-01-21~\cite{gemini}, DeepSeek-R1 ~\cite{guo2025deepseek}, and OpenAI o1-2024-12-17 \cite{openai}.

\textbf{Gemini-2.0-Flash-Thinking-Exp-01-21} ~\cite{gemini} is an experimental reasoning model released by Google in January 2025. The model adopts a unique ``slow thinking'' mode, which shows excellent performance in complex tasks such as mathematics and programming by breaking down user prompts step by step and generating a traceable complete CoT. In addition, the model supports context windows of up to 1 million tokens and has multimodal reasoning capabilities, making it more advantageous when processing large-scale and multi-type data.

\textbf{DeepSeek-R1}~\cite{guo2025deepseek} is an open-source reasoning model launched by DeepSeek in January 2025. The model uses a Mixture of Experts (MoE) architecture~\cite{jacobs1991adaptive} with 671B parameters, 37B per token activation, and 128K context windows, enhancing efficiency for large-scale data processing. DeepSeek-R1 uses large-scale reinforcement learning to boost reasoning, excelling at math, code, and reasoning tasks.

\textbf{OpenAI o1-2024-12-17}~\cite{openai} is a closed-source model released by OpenAI at the end of 2024. As a new post-training version of the ChatGPT model, it is known for its excellent reasoning ability. The model performs well in benchmarks such as mathematics, science, and coding, fully demonstrating its strength in complex reasoning tasks.


\subsubsection{Datasets}
To analyze the quality of the CoT generated by LLMs, we selected two popular code benchmarks in real-world software development scenarios: CoderEval~\cite {Zhang_2024} and SWE-bench~\cite{jimenez2023swe}.

\textbf{CoderEval} consists of 230 functions from 43 Python projects and 230 methods from 10 Java projects, which are all selected from high-star open source projects of various domains. Each task includes the docstring, signature, reference solution code, and unit tests to evaluate generated code correctness. This study primarily focuses on the 230 Python tasks.

\textbf{SWE-bench} is designed to evaluate language models' performance in software engineering tasks, specifically focusing on automatically resolving GitHub issues.
It contains 2,294 instances from open-source projects, such as GitHub repositories, where each instance includes an issue description, the relevant code snippet, and the corresponding fix patch. The benchmark requires models to generate accurate solutions for new problems. Additionally, it is available in three versions: Full, Lite, and Verified. 
The Full version comprises a diverse set of issue types, including New Features, Bug Fixing, and Enhancements, among others. New Features typically require the development of entirely new code segments or functions to address the issue. 
As the primary focus of this study is on code generation tasks, we selected a subset of 111 instances related to New Features from the Full version, termed \textit{SWE-bench-NF}, for detailed analysis. 




To ensure that the model achieves optimal inference performance, we strictly adhere to the preset parameter configurations.
Furthermore, since the reasoning model can generate reasoning steps and final code based on user queries, we directly input the requirements and contexts from the original benchmark into three different LLMs to generate CoTs and final codes.
Subsequently, we evaluated the correctness of generated code in accordance with the assessment protocols outlined in the original benchmark. 
This process yields 1,023 code samples with associated CoTs, comprising 690 from CoderEval and 333 from SWE-bench-NF.

\subsubsection{Construction of CoT quality taxonomy}

During CoT quality classification construction, we hypothesize that incorrect code exhibits stronger correlation with low-quality CoTs.
Based on this, this section focuses on 813 code samples that failed the test cases and their corresponding CoTs.


First, we recruit four experts in the evaluation, 
each with over ten years of experience in Java and Python development, engaged in focused discussions and in-depth analysis on the potential factors behind the quality issues of CoT generated by LLMs. 
During this process, we randomly selected 262 samples from 813 failed CoT samples (with a confidence level of 95\% and a margin of error of ±5\%) and employed the Open Card Sorting method to explore the underlying factors systematically.
Each sample was treated as an individual ``card'', with each card representing a CoT sample.
The four experts were divided into two independent groups to evaluate the poor-quality CoT samples. Each group analyzed every reasoning step based on the requirement description, reference code solutions, and execution results (such as error messages), aiming to identify errors and their root causes. Any discrepancies in annotations were thoroughly discussed among the experts until a consensus was reached. To verify the consistency of the classification, 
we calculated the Cohen's Kappa, which was found to be 82\%, indicating perfect agreement and highlighting the reliability of our procedure.
Moreover, the intermediate reasoning steps for each requirement may involve multiple factors contributing to the poor quality of the CoT. To delve deeper into the quality issues, experts were asked to assess which factor had the greatest impact on the correctness of the code in each sample. If multiple factors were at play, samples were allowed to be labeled with multiple tags. 
Through this process, the experts created a codebook that provided a structured classification framework, forming the basis for subsequent analysis.

Second, using the codebook, the remaining 551 samples were annotated by four participants  (including one PhD student, two graduate students, and one undergraduate student).
All participants had experience in software development research or open-source projects. 
Each sample was independently annotated by two participants to eliminate subjective bias, following the same procedure as the pilot analysis to refine and expand the taxonomy.
If a new category not covered by the predefined ones emerged, annotators were required to describe the reasons for the poor CoT quality to facilitate further discussion for establishing new classifications. 
We calculated the Cohen's Kappa scores for each pair of participants annotating the same set of samples, with all average scores above 0.7. For any discrepancies in the annotations, the corresponding participants and the three authors held meetings to discuss the differences until a consensus was reached.

\begin{figure*}[t]
\centering
\setlength{\abovecaptionskip}{0.1cm}
\includegraphics[width=0.85\textwidth]{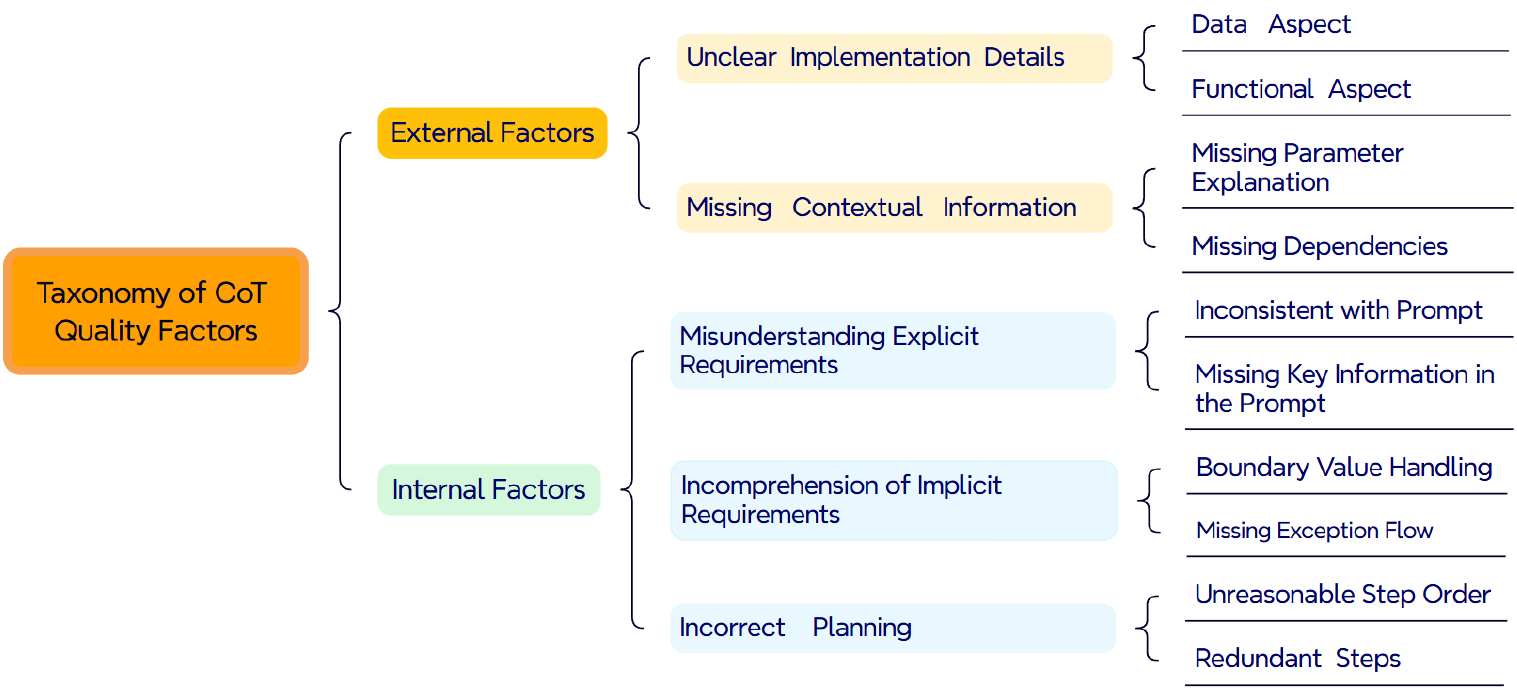}
\caption{Taxonomy of factors influencing CoT quality}
\label{fig: class}
\vspace{-0.35cm}
\end{figure*}

Finally, we developed a classification framework to analyze the factors influencing the quality of CoT, as depicted in Figure~\ref{fig: class}. 
This framework is categorized into two primary dimensions: \textbf{External Factors} and \textbf{Internal Factors}. 
External factors primarily encompass Unclear Implementation Details and Missing Contextual Information. 
Unclear Implementation Details refer to ambiguities at the data level—such as vague specifications of data structures or functionalities—and insufficiently detailed descriptions of function implementations, where the intended functionality lacks clear articulation.
Missing Contextual Information refers to absent critical parameters and dependencies, appearing as inaccurate or incomplete references that impair full understanding.

In contrast, internal factors are evaluated across three key aspects: Misunderstanding Explicit Requirements, Incomprehension of Implicit Requirements, and Incorrect Planning. 
Misunderstanding Explicit Requirements arises when the CoT conflicts with the provided prompt or omits critical information explicitly stated in the prompt, leading to misalignment with the intended objectives. 
Incomprehension of Implicit Requirements is characterized by incomplete coverage within the CoT’s logical flow, such as the failure to address boundary conditions or incorporate exception handling, which are essential for robustness. 
Lastly, Incorrect Planning refers to flaws in the sequencing of steps within the CoT, including illogical orderings or the inclusion of redundant steps that detract from efficiency and coherence.


\subsection{External Factors}
\label{sec: external}

\subsubsection{\textbf{Unclear Implementation Details (UID)}}

Unclear implementation details primarily manifest in two aspects: data and functionality. In terms of data, this mainly refers to insufficiently defined data structures or data types in the requirements, which leads to inconsistent data processing logic during the model's reasoning process. Regarding functionality, the issue primarily focuses on unclear descriptions of specific functions, such as failing to specify whether a sorting operation should be in ascending or descending order, or not providing a matching pattern in regular expression operations. 
Regardless of whether it is ambiguity in data definitions or a lack of functional descriptions, both can lead to erroneous reasoning by the model~\cite{10.1145/3660810}.

\begin{figure}
    \centering
    \setlength{\abovecaptionskip}{0.1cm}
    \includegraphics[width=0.45\textwidth,height=8cm]{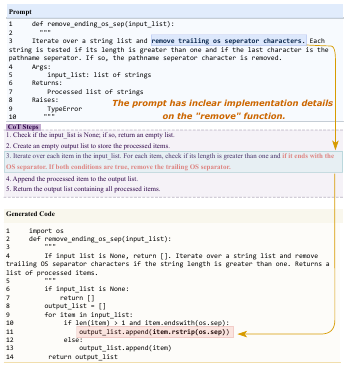}
    \caption{Example of Unclear Implementation Details } 
    \label{unclear detailsfig}
    \vspace{-0.65cm}
\end{figure}

An example of ``Unclear implementation details'' is shown in Figure~\ref{unclear detailsfig}.
In the prompt, the requirement involves removing the trailing operating system path separator (os.sep) from a list of strings, with specific constraints provided. If a string’s length is greater than 1 and ends with os.sep, the trailing os.sep should be removed; otherwise, the string remains unchanged. At first glance, this requirement appears straightforward. However, upon closer examination of specific scenarios, a critical issue emerges that is not clearly addressed: how should multiple consecutive os.sep characters at the end of a string be handled? For example, consider an input\_list like [`a', `b', `c' + os.sep * 2], where there are two os.sep characters. According to the prompt’s description, we know it requires ``removing the trailing os.sep.'' Assumption 1: Remove all trailing os.sep characters (e.g., using Python’s rstrip(os.sep)), resulting in [`a', `b', `c']. Assumption 2: The LLM removes only one os.sep, resulting in [`a', `b', `c\textbackslash\textbackslash']. 
As shown in step 3 of Figure~\ref{unclear detailsfig}. 
If the LLM adopts Assumption 1, but the task actually expects Assumption 2 (based on test cases), the implementation will be incorrect. Without explicit prompt clarification, the LLM might remove all consecutive os.sep characters during reasoning and implementation, causing a mismatch with the intended result.

\subsubsection{\textbf{Missing Contextual Information (MCI)}}

This type of external factor refers to the absence of essential contextual information within the requirements that is essential for completing the task.                                
Based on our preliminary analysis, this mainly includes missing parameter explanations and missing dependencies.

\begin{figure}[t]
    \centering
    \setlength{\abovecaptionskip}{0.1cm}
    \includegraphics[width=0.4\textwidth,height=6.5cm]{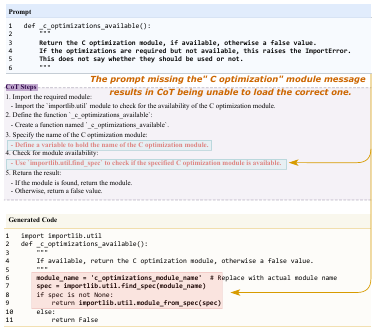}
    \caption{Example of Missing Dependencies } 
    \label{context fig}
    \vspace{-0.65cm}
\end{figure}

Missing dependencies refer to external functions, methods, or attributes that are required to be called during the process of function implementation but are not included. However, the requirements did not specify this clearly, resulting in errors in the generated code.
These types of errors frequently occur in the CoderEval dataset. For example, as illustrated in Figure~\ref{context fig}, the primary function of this task is to check whether a certain C-language optimization module is available and, based on the result, take the appropriate action.
If the module is available, it should return the module; otherwise, it should return ``\textit{false}''.
If the C optimization module is mandatory (i.e., the program's execution depends on it) but unavailable, it should raise an ``\textit{ImportError}'' exception.
From the reference code, it is clear that the function needs to check for a specific C-language module.
However, the prompt does not specify which external C optimization module needs to be checked.
This ambiguity leads the LLMs to generate code using a placeholder variable like ``\textit{c\_optimizations\_module\_name}'' instead of specifying the actual module name. 
While this approach might seem reasonable, it fails the test when directly executed. 
To avoid such issues, when it is expected to check for a specific module, the prompt should explicitly mention the required external module to ensure that the generated code correctly references the intended module.

\begin{tcolorbox}[colback=gray!20, colframe=lightgray, boxrule=1pt, arc=5pt,left=5pt,right=5pt,top=5pt,bottom=5pt,before skip=12pt]
\textbf{Finding 1:} Among external factors, 55.6\% lacked contextual information, while 44.4\% of the requirements had unclear descriptions of function implementation details. 
This highlights the importance of providing precise contextual information and clearly defined requirements for coding tasks.

\end{tcolorbox}

\subsection{Internal factors}
\label{sec: internal factors}

\subsubsection{\textbf{Misunderstanding Explicit Requirements (MER)}}

Explicit requirement understanding refers to content that is clearly described in the requirements, but the model fails to capture or correctly interpret a key piece of information during the reasoning process. This is mainly reflected in the CoT's failure to capture the key information in the prompt and the CoT being inconsistent with the prompt content.

\begin{figure}[t]
    \centering
    \setlength{\abovecaptionskip}{0.1cm}
    \includegraphics[width=0.4\textwidth,height=6cm]{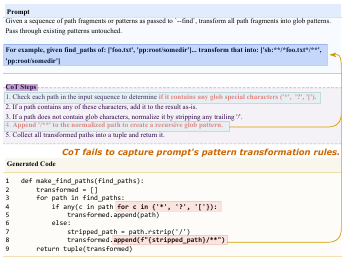}
    \caption{Example of CoT Inconsistent with Prompt } 
    \label{CoT Inconsistent fig}
    \vspace{-0.6cm}
\end{figure}

Figure~\ref{CoT Inconsistent fig} shows an example of this category. The purpose of this task is to convert a sequence of path fragments or patterns (provided as input to \--{find}) into glob patterns while preserving existing patterns unchanged. For example, the input [`foo.txt', `pp:root/somedir'] should be transformed into [`sh:**/*foo.txt*/**', `pp:root/somedir']. Here, foo.txt is treated as a path fragment and converted into the complex glob pattern sh:**/*foo.txt*/**, whereas pp:root/somedir is considered an existing pattern and remains unchanged. This indicates that the core of the task lies in accurately distinguishing between ``path fragments'' and ``existing patterns'' and applying specific glob conversion rules to path fragments. 
During implementation, the LLM first checks each path for the presence of glob special characters (such as *, ?, or [ ). If these characters are present, the path is treated as an existing pattern and preserved as is; if absent, it is regarded as a path fragment, normalized (by removing trailing /), and appended with /**. Finally, all processed results are collected and returned as a tuple. While these steps appear reasonable, there is a significant inconsistency with the original requirements. The issues primarily manifest in two areas: 
\textbf{Inadequate Pattern Recognition.} The CoT relies solely on the presence of glob special characters to determine whether a path is an “existing pattern.” In the example, pp:root/somedir contains no *, ?, or [, and according to the CoT logic, it would be treated as a path fragment and converted to pp:root/somedir/**. However, the requirement explicitly states that it should remain unchanged (as pp:root/somedir), indicating that pp:root/somedir should be recognized as an existing pattern. The LLM overlooks the requirement to preserve existing patterns unchanged during its reasoning process.
\textbf{Deviation in Conversion Rules.} For path fragments, the CoT’s conversion rule simply appends /** to the path, such as transforming foo.txt into foo.txt/**. However, in the requirement’s example, foo.txt is converted into sh:**/*foo.txt*/**, a more complex glob pattern that includes the prefix sh: and nested * structures. The CoT’s rule fails to produce this result, demonstrating that its conversion logic does not fully capture the complete requirements of the task.

\subsubsection{\textbf{Incomprehension of Implicit Requirements (IIR)}}
In certain tasks, there exist implicit requirements that are not explicitly stated. The model fails to recognize these requirements during the reasoning process, resulting in incomplete logical flows, such as inadequate handling of boundary values or missing exception flows.

\begin{figure}
    \centering
    \setlength{\abovecaptionskip}{0.1cm}
    \includegraphics[width=0.45\textwidth,height=6.5cm]{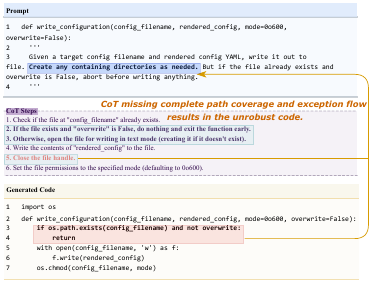}
    \caption{Example of Implicit Requirement Misinterpretation} 
    \label{ImplicitRequirement}
    \vspace{-0.6cm}
\end{figure}

Figure~\ref{ImplicitRequirement} shows an example of this category. The task primarily involves writing the rendered YAML content (rendered\_config) to a specified target file (config\_filename). According to the prompt’s requirement, “Create any containing directories as needed,” the program must automatically create the directories containing the target file if they do not exist. Additionally, if the target file already exists and the overwrite parameter is set to False, the program should abort before writing to avoid overwriting the existing file.
As shown in Figure~\ref{ImplicitRequirement}, the LLM's reasoning process unfolds as follows. Steps 1-2: First, check whether the target file exists and whether the overwrite parameter is False. If the file exists and overwriting is not allowed, exit early to prevent subsequent write operations. Steps 3-5: If writing is permitted, follow a standard file operation workflow—open the file, write the rendered\_config content, and close the file. Step 6: Finally, set the file’s permissions.
This process appears logically complete, but the LLM's reasoning overlooks a critical requirement from the prompt regarding directory creation. 
Specifically, the CoT lacks a step to check and create the directory where the target file is located before writing the file.
If the directories do not exist, attempting to write the file directly will result in an operation failure. Therefore, the correct workflow should be: first, check if the file exists and decide whether to proceed based on the overwrite parameter; then, create the necessary directories; and finally, perform the file write operation. Only this approach fully satisfies the prompt’s requirements.

Furthermore, from the perspectives of security and robustness, when the target file exists and overwrite is False, the program should raise a FileExistsError exception to explicitly alert the user and prevent accidental overwriting of important files. During directory creation, exception handling (e.g., FileExistsError, FileNotFoundError) should be added to manage edge cases, ensuring stable directory creation.

\subsubsection{\textbf{Incorrect Planning (ICP)}}
This type primarily considers situations where the reasoning steps may have unreasonable sequences or redundant steps, causing the CoT to deviate from the intended goal.

\begin{figure}[t]
    \centering   
    \setlength{\abovecaptionskip}{0.1cm}
    \includegraphics[width=0.45\textwidth,height=7.5cm]{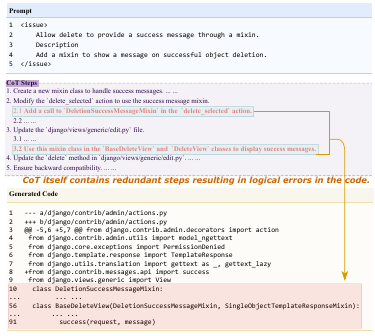}
    \caption{Example of Incorrect Planning} 
    \label{incorrectPlanning fig}
    \vspace{-0.4cm} 
\end{figure}

Figure~\ref{incorrectPlanning fig} shows an example of the Incorrect Planning`` category. 
This task primarily involves adding a mixin to Django’s deletion operations (e.g., DeleteView) to display a success message after an object is successfully deleted. This involves crafting a reusable mixin to trigger and show a success message after deletion, notifying the user of completion.
To implement this functionality, the LLM first creates a mixin class named `DeletionSuccessMessageMixin' and includes a `success\_message' attribute to store a custom message. Second, it considers applying this mixin to the `delete\_selected' action in Django admin to support bulk deletion messages. Third, it suggests integrating the mixin into `BaseDeleteView' and overriding the `delete' method to ensure the message is displayed post-deletion while avoiding the redundant application of the mixin in `DeleteView'. Finally, it emphasizes ensuring backward compatibility to maintain the stability of existing code.
During this reasoning process, two main issues arise. First, modifying `delete\_selected' is unnecessary. This requirement focuses on Django’s generic views, particularly deletion-related ones, not admin functionalities. Second, applying the mixin in `DeleteView' introduces redundancy. Since `DeleteView' inherits from `BaseDeleteView', integrating the mixin into `BaseDeleteView' alone suffices. Thus, step 3.2—reapplying the mixin in `DeleteView'—is redundant.

\begin{tcolorbox}[colback=gray!20, colframe=lightgray, boxrule=1pt, arc=5pt,left=5pt,right=5pt,top=5pt,bottom=5pt,before skip=12pt]
\textbf{Finding 2:} Among internal factors, a significant majority, 59.7\%, of the poor CoT instances are caused by the model's failure to comprehend implicit requirements, while a substantial 35.8\% stem from the model's inability to understand explicit requirements.
Enhancing the model's capability to uncover implicit requirements is essential for improving the quality of CoT.

\end{tcolorbox}

\subsection{Distribution on CoT Quality Factors}

\label{section: Overview of CoT Quality}





This study mainly analyzes the quality of CoT generated by DeepSeek-R1, o1-2024-12-17, and Gemini-2.0-Flash-Thinking on CoderEval and SWE-bench-NF.
First, we evaluate the pass@1 metric on both benchmarks and comprehensively assess the quality of CoT.
As shown in Table~\ref{tab:tj-1}, DeepSeek-R1 achieves the highest Pass@1 scores, recording 30.43\% on CoderEval and 6.31\% on SWE-bench-NF, outperforming the o1 model (28.70\% on CoderEval and 4.50\% on SWE-bench-NF) and Gemini-2.0-Flash-Thinking(25.65\% on CoderEval and 2.70\% on SWE-bench-NF).
\textbf{Additionally, across the 1023 CoT-code pairs generated for both benchmarks, 76.4\% (782/1023) of the CoTs exhibited quality issues.} Specifically, in CoderEval, 73.6\% (508/690) of the CoTs were problematic, while in SWE-bench-NF, this proportion increased to 82.3\% (274/333).

In CoderEval, DeepSeek-R1 achieves a Pass@1 score of 30.43\%, which is better than the 28.70\% of the o1-2024-12-17 and the 25.65\% of Gemini-2.0-Flash-Thinking.
However, an analysis of CoT quality revealed that, among the 690 CoT-code pairs, 508 CoTs were incorrect, leaving only 182 correct. 
The quality issues were particularly evident in the Gemini-2.0-Flash-Thinking and o1-2024-12-17 models. For Gemini-2.0-Thinking, out of 59 codes that passed the test cases, 18 had incorrect corresponding CoTs, while among the 171 codes that failed, 8 had correct CoTs. 
Similarly, for the o1-2024-12-17 model, 7 out of 66 passing codes had incorrect CoTs, and 4 out of 164 failing codes had correct CoTs.
In SWE-bench-NF, DeepSeek-R1 maintained its lead with a Pass@1 of 6.31\%, compared to 4.50\% for the o1 model and 2.70\% for Gemini-2.0-Flash-Thinking. 
Examination of the CoT quality showed that, out of 333 CoT-code pairs, 274 CoTs were incorrect, with only 59 being correct. 
A detailed analysis indicated that, for DeepSeek-R1, 14 out of 104 codes that failed the test cases had correct corresponding CoTs. For Gemini-2.0-Flash-Thinking, among 108 failing codes, 16 had correct CoTs, while 92 had incorrect CoTs. Likewise, for the o1-2024-12-17 model, 14 out of 106 failing codes had correct CoTs, with 92 having incorrect CoTs.

\begin{table}[htbp]
\scriptsize  
\setlength{\tabcolsep}{3pt}  
\caption{Comparison of Pass@1, Pass/Fail Test Cases, and CoT Correct/Incorrect for LLMs on CoderEval and SWE-bench-NF}
\begin{tabular}{>{\centering\arraybackslash}p{1.1cm} >{\centering\arraybackslash}p{1.5cm} >{\centering\arraybackslash}p{0.7cm} >{\centering\arraybackslash}p{0.7cm} >{\centering\arraybackslash}p{0.7cm} >{\centering\arraybackslash}p{1.1cm} >{\centering\arraybackslash}p{1.1cm}}
\toprule
\multirow{2}{*}{\textbf{Datasets}} & 
\multirow{2}{*}{\textbf{Model}} & 
\multirow{2}{*}{\textbf{Pass@1}} & 
\multicolumn{2}{c}{\textbf{Pass/Fail Test Case}} & 
\multirow{2}{*}{\begin{tabular}{c} 
                  \textbf{CoT} \\ 
                  \textbf{Correct} 
                \end{tabular}} & 
\multirow{2}{*}{\begin{tabular}{c} 
                  \textbf{CoT} \\ 
                  \textbf{Incorrect} 
                \end{tabular}} \\
                

\cmidrule{4-5}  
& & & \textbf{Pass/Fail} & \textbf{Num} & & \\
\midrule
\multirow{7}{1.1cm}{\centering CoderEval} & \multirow{2}{*}{DeepSeek-R1} & \multirow{2}{*}{30.43\%} & Pass & 70 & 70 & 0 \\
& & & Fail & 160 & 0 & 160 \\ 
\cmidrule{2-7}
& \multirow{2}{*}{\begin{tabular}[c]{@{}c@{}}Gemini-2.0\end{tabular}} & \multirow{2}{*}{25.65\%} & Pass & 59 & 41 & 18 \\
& & & Fail & 171 & 8 & 163 \\
\cmidrule{2-7}
& \multirow{2}{*}{o1-2024-12-17} & \multirow{2}{*}{28.70\%} & Pass & 66 & 59 & 7 \\
& & & Fail & 164 & 4 & 160 \\
\cmidrule{2-7}
& \multicolumn{1}{c}{\multirow{1}{*}{\centering Total}} & --& -- & \textbf{690} & \textbf{182} & \textbf{508} \\
\midrule
\multirow{7}{1.1cm}{\centering SWE-bench-NF} & \multirow{2}{*}{DeepSeek-R1} & \multirow{2}{*}{6.31\%} & Pass & 7 & 7 & 0 \\
& & & Fail & 104 & 14 & 90 \\
\cmidrule{2-7}
& \multirow{2}{*}{\begin{tabular}[c]{@{}c@{}}Gemini-2.0\end{tabular}} & \multirow{2}{*}{2.70\%} & Pass & 3 & 3 & 0 \\
& & & Fail & 108 & 16 & 92 \\
\cmidrule{2-7}
& \multirow{2}{*}{o1-2024-12-17} & \multirow{2}{*}{4.50\%} & Pass & 5 & 5 & 0 \\
& & & Fail & 106 & 14 & 92 \\
\cmidrule{2-7}
& \multicolumn{1}{c}{\multirow{1}{*}{\centering Total}} & -- & -- & \textbf{333} & \textbf{59} & \textbf{274} \\
\midrule
\multicolumn{2}{c}{\centering Total} & -- & -- & \textbf{1023} & \textbf{241} & \textbf{782} \\
\bottomrule
\end{tabular}
\label{tab:tj-1}
\end{table}

\begin{figure}[htbp]
    \centering
    \begin{subfigure}[t]{0.45\linewidth}
        \centering
        \includegraphics[width=\linewidth]{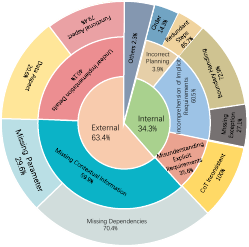}
        \caption{CoderEval}
        \label{tx-1}
    \end{subfigure}
    \hfill
    \begin{subfigure}[t]{0.45\linewidth} 
        \centering   
        \includegraphics[width=\linewidth]{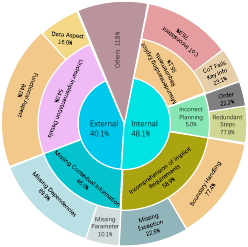}
        \caption{SWE-bench-NF}
        \label{tx-2}
    \end{subfigure}
    \caption{Distribution of CoT Quality Factors on CoderEval and SWE-bench-NF}
    \label{fig:two_pictures}
    \vspace{-0.3cm} 
\end{figure}

Second, we systematically analyze the distribution of low-quality CoT factors across both benchmarks (CoderEval and SWE-bench-NF).
In CoderEval, as shown in Figure~\ref {tx-1}, external factors are the primary drivers of CoT quality issues, accounting for 63.4\%.
These are primarily driven by missing contextual information (particularly dependencies) and unclear implementation details (especially functional descriptions), which constitute 42.2\% and 31.8\% of the external factors, respectively. Internal factors predominantly arise from incomplete consideration of boundary cases (44.1\%) and inconsistencies between the CoT and the prompt (35.6\%).
In contrast, as shown in Figure~\ref {tx-2}, in the \textit{SWE-bench-NF}, the factors impacting CoT quality are more evenly distributed, with internal factors contributing 48.1\% and external factors 40.1\%. 
Internal factors account for 48.1\%, driven primarily by incomplete handling of boundary cases (45.6\%) and CoT prompt inconsistencies (27.8\%), while external factors contribute 40.1\%, resulting mainly from ambiguous functional-level details (45.3\%) and inaccurate contextual dependencies (41.3\%). 

Finally, our analysis reveals that external factors constitute 53.6\% of the total, while internal factors make up the remaining 40.1\%.
We also investigate the distribution of these 649 factors across LLMs.
Table~\ref{tab:merged} presents the distribution of factors across various LLMs in CoderEval and SWE-bench-NF.
A darker color indicates a higher degree to which the factor contributes to CoT quality issues.
In the analysis of CoderEval, among external factors, missing dependencies most frequently lead to erroneous CoT generation across all three models, followed by unclear implementation details (functional aspect). 
Notably, compared to DeepSeek-R1 (17.2\%) and o1 (18.8\%), Gemini-2.0-Flash-Thinking (10.2\%) appears to handle boundary cases more effectively.
In SWE-bench-NF, among external factors, unclear functional descriptions in requirements are the most likely to lead to Gemini-2.0-Flash-Thinking generating erroneous CoT.
Conversely, inaccurate contextual dependencies more significantly degrade the CoT quality for DeepSeek-R1 and o1. 
Regarding internal factors, the CoT produced by DeepSeek-R1 and o1 is more susceptible to incomplete consideration of boundary cases compared to Gemini-2.0-Flash-Thinking. 
However, the proportion of CoT inconsistencies with the prompt remains nearly identical across all three models.

Furthermore, to examine the impact of CoTs on final code correctness, we analyze the quality of CoTs associated with passing code samples (210 instances).
\textbf{We found that 11.9\% of the CoTs had quality issues, even though the corresponding code was correct. }
This may be attributed to the strong error tolerance of LLMs. To this end, we conducted a preliminary analysis on the relationship between CoT and code generation. The results show that even when the CoT is correct, there is approximately an 18.5\% probability of generating incorrect code. \textbf{Conversely, even when the CoT is incorrect, there is a 3.1\% probability of generating correct code}. These findings indicate that LLMs do not strictly adhere to their own CoT steps when generating code.
\begin{tcolorbox}[colback=gray!20, colframe=lightgray, boxrule=1pt, arc=5pt,left=5pt,right=5pt,top=5pt,bottom=5pt]
\textbf{Finding 3:} In the generated CoT-Code pairs, 76.4\% of the CoTs have quality issues. Interestingly, we found that 11.9\% of the CoTs exhibit quality issues, despite the corresponding code being correct. Furthermore, even with a correct CoT, there remains an 18.5\% probability of generating erroneous code. Conversely, even with an incorrect CoT, there is a 3.1\% probability of producing correct code.
\end{tcolorbox}

\begin{table}[t!]
\setlength{\abovecaptionskip}{0.1cm}
\caption{Distribution of Factors Across Different LLMs}
\tiny 
\renewcommand{\arraystretch}{1.2} 
\setlength{\tabcolsep}{3pt} 

\begin{tabular}{| 
    >{\centering}m{0.8cm}| 
    >{\centering}m{0.6cm}| 
    >{\centering}m{1.7cm}| 
    *{6}{>{\centering\arraybackslash}m{0.6cm}|}} 
\hline
\multicolumn{3}{|>{\centering\arraybackslash}m{3.2cm}|}{\textbf{Factors}\rule{0pt}{12pt}}
    & \multicolumn{3}{c|}{\textbf{CoderEval}} 
    & \multicolumn{3}{c|}{\textbf{SWE-bench-NF}} \\
\cline{4-9}
\multicolumn{3}{|c|}{} 
    & \textbf{D-R1} & \textbf{G-2.0} & \textbf{o1} 
    & \textbf{D-R1} & \textbf{G-2.0} & \textbf{o1} \\ 
\hline

\multirow{5}{0.9cm}{\centering \textbf{External}} 
    & \multirow{2}{0.6cm}{\centering UID} 
    & Data Aspect 
    & \cellcolor[HTML]{E8F4E9}1.9\% 
    & \cellcolor[HTML]{BFE3CB}8.4\% 
    & \cellcolor[HTML]{D7EDE0}5.5\% 
    & \cellcolor[HTML]{E0F1E8}4.8\% 
    & \cellcolor[HTML]{D7F0E5}5.1\% 
    & \cellcolor[HTML]{EDF6F3}1.9\% \\ 
\cline{3-9} 
& & Functional Aspect 
    & \cellcolor[HTML]{6DC68B}18.5\% 
    & \cellcolor[HTML]{63BE7B}22.9\% 
    & \cellcolor[HTML]{6FC78D}20.4\% 
    & \cellcolor[HTML]{B0DFBF}14.4\% 
    & \cellcolor[HTML]{63BE7B}29.7\% 
    & \cellcolor[HTML]{9AD8A4}16.7\% \\ 
\cline{2-9} 
& \multirow{2}{0.6cm}{\centering MCI} 
    & Missing Parameter 
    & \cellcolor[HTML]{8DD4A3}12.7\% 
    & \cellcolor[HTML]{98D4A9}14.5\% 
    & \cellcolor[HTML]{CDE9D7}7.7\% 
    & \cellcolor[HTML]{E8F4EF}2.9\% 
    & \cellcolor[HTML]{EDF6F3}0.8\% 
    & \cellcolor[HTML]{E8F5E9}2.8\% \\ 
\cline{3-9} 
& & Missing Dependencies 
    & \cellcolor[HTML]{63BE7B}27.4\% 
    & \cellcolor[HTML]{63BE7B}22.9\% 
    & \cellcolor[HTML]{63BE7B}31.5\% 
    & \cellcolor[HTML]{8FCE9F}19.2\% 
    & \cellcolor[HTML]{9BD4A2}16.1\% 
    & \cellcolor[HTML]{7FC88D}21.3\% \\ 
\cline{2-9} 
& \multicolumn{2}{c|}{Total} 
    & 60.51\% & 68.67\% & 65.19\% 
    & 41.3\% & 51.7\% & 42.6\% \\ 
\hline

\multirow{7}{0.9cm}{\centering \textbf{Internal}} 
    & \multirow{2}{0.6cm}{\centering MER} 
    & CoT Inconsistent 
    & \cellcolor[HTML]{8DD4A3}12.7\% 
    & \cellcolor[HTML]{A8DAB7}12.0\% 
    & \cellcolor[HTML]{B7E0C4}12.7\% 
    & \cellcolor[HTML]{A3D8B1}15.4\% 
    & \cellcolor[HTML]{A0D6A6}15.3\% 
    & \cellcolor[HTML]{A5DCAC}14.8\% \\ 
\cline{3-9} 
& & CoT Fails Key Info 
    & 0 
    & 0 
    & 0 
    & \cellcolor[HTML]{EDF6F3}1.9\% 
    & \cellcolor[HTML]{D3EFE3}5.9\% 
    & \cellcolor[HTML]{D6F0D5}5.6\% \\ 
\cline{2-9} 
& \multirow{2}{0.6cm}{\centering IIR} 
    & Boundary Handling 
    & \cellcolor[HTML]{9CD5AC}17.2\% 
    & \cellcolor[HTML]{B4DFC1}10.2\% 
    & \cellcolor[HTML]{6FC78D}18.8\% 
    & \cellcolor[HTML]{63BE7B}28.8\% 
    & \cellcolor[HTML]{8ACE94}19.5\% 
    & \cellcolor[HTML]{63BE7B}26.9\% \\ 
\cline{3-9} 
& & Missing Exception 
    & \cellcolor[HTML]{B4E3C0}8.3\% 
    & \cellcolor[HTML]{C3E5CF}7.8\% 
    & \cellcolor[HTML]{E8F4EF}1.7\% 
    & \cellcolor[HTML]{D0ECDC}8.7\% 
    & \cellcolor[HTML]{C9EDDF}6.8\% 
    & \cellcolor[HTML]{CFEDCF}6.5\% \\ 

\cline{2-9} 
& \multirow{2}{0.6cm}{\centering ICP} 
    & Unreasonable Order 
    & 0 
    & 0 
    & \cellcolor[HTML]{EDF6F3}0.6\% 
    & 0 
    & 0 
    & 0.9\%  \\ 

\cline{3-9} 
& & Redundant Steps 
    & \cellcolor[HTML]{EDF6F3}1.3\% 
    & \cellcolor[HTML]{EDF6F3}1.2\% 
    & \cellcolor[HTML]{EBF5F1}1.1\% 
    & \cellcolor[HTML]{E8F4EF}2.9\% 
    & \cellcolor[HTML]{EDF6F3}0.8\% 
    & \cellcolor[HTML]{E8F5E9}2.8\% \\ 
\cline{2-9} 
& \multicolumn{2}{c|}{Total} 
    & 39.49\% & 31.33\% & 34.81\% 
    & 58.7\% & 48.3\% & 57.4\% \\ 
\hline


\end{tabular}
\label{tab:merged}
\vspace{-0.4cm}
\end{table}

\section{Automatic Detection via Multi-Agent Debate}

\subsection{Study Design}

The findings in the previous section reveal that low-quality CoTs can significantly degrade the final code quality, which urgently requires an automated detection mechanism.
Therefore, we introduce a Multi-Agent Debate (MAD)~\cite{liang2023encouraging} framework, which leverages multi-party debate perspectives as feedback to identify low-quality CoTs.
This framework simulates human debate dynamics, enabling multiple agents to negotiate specific issues toward consensus.
Each agent embodies a distinct perspective or strategy; through structured argumentation, rebuttals, and collaborative deliberation, they iteratively explore viable solutions.
We assign three specialized roles(\textbf{Verifier}, \textbf{defender}, and \textbf{Arbiter}) to automate detection.
The verifier is responsible for questioning the potential quality problems of the CoT generated by the defender. 
The defender's role is to maintain the correctness of the original CoT by presenting evidence and logical reasoning to refute the verifier's doubts.
The arbitrator performs a comprehensive analysis of the dispute's focal points, evaluates the rationality of the arguments, and makes the final decision.
Within this framework, the Defender, Verifier, and Arbiter roles are instantiated from the reasoning model in Section 3.
Specifically, \ding{182}\textbf{MAD\_DGO} means DeepSeek-R1 is the defender, Gemini-2.0-Flash-Thinking is the arbitrator, and o1 is the verifier; \ding{183}\textbf{MAD\_GDO} means Gemini-2.0-Flash-Thinking is the defender, DeepSeek-R1 is the arbitrator, and o1 is the verifier. \ding{184}\textbf{MAD\_OGD} means o1 defender, Gemini-2.0-Flash-Thinking arbitrator, and DeepSeek-R1 verifier.

The debate process consists of three rounds, with the provision to terminate early if a conclusion is reached before all rounds are completed.
In addition, we employ Recall, Precision, and F1-Score as evaluation metrics to evaluate the ability of the MAD method to detect erroneous CoTs and the accuracy of its detection, respectively.

\subsection{Results}

Table~\ref{tab:adjusted_table} comparatively evaluates multiple methods using Recall, Precision, and F1 metrics across CoderEval and SWE-bench-NF. 
Methods sharing the same color scheme denote comparative pairs in our analysis. 
For example, DeepSeek-R1 (highlighted in green) compares directly with MAD\_DGO (similarly green), while Gemini-2.0-Flash-Thinking (marked in black) evaluates against MAD\_GDO (matching black). 
Overall, MAD-series variants consistently outperform single LLM detection approaches on both datasets.
On CoderEval, MAD-series methods (specifically MAD\_DGO and MAD\_OGD) exhibit superior recall (42.5\% and 43.9\% respectively), significantly exceeding alternatives. 
However, their precision is notably lower, with MAD\_DGO achieving 12.50\% and MAD\_OGD only 5.49\%.
Single-LLM methods, DeepSeek-R1 delivers balanced performance with a recall of 8.75\%, precision of 6.25\%, and F1-score of 7.29\%. In contrast, Gemini and o1 exhibit weaker performance across all metrics, suggesting limitations in their error detection capabilities and precision.

On SWE-bench-NF, all methods demonstrate improved performance compared to CoderEval. 
The MAD-series variants, MAD\_DGO and MAD\_OGD, demonstrate superior recall performance on CoderEval, achieving recall of 46.36\% and 48.54\%, respectively.
Notably, MAD\_OGD records the highest recall but suffers from a precision of only 4.85\%, resulting in an F1-score of 8.83\%, which highlights a significant performance imbalance. 
Single LLM methods, DeepSeek-R1, Gemini-2.0-Flash-Thinking, and o1 show enhanced results. DeepSeek-R1 stands out with a recall of 23.64\%, precision of 20.91\%, and F1-score of 22.19\%, reflecting strong overall performance and adaptability to the SWE-bench-NF task.

\begin{table}[t]
\caption{Comparison of MAD and single LLM Methods in Recall, Precision, and F1-Score}
\label{tab:metrics_comparison}
\centering
\scriptsize
\setlength{\tabcolsep}{3.5pt}
\renewcommand{\arraystretch}{1.2}
\begin{tabular}{@{} >{\centering\arraybackslash}m{1.5cm} >{\centering\arraybackslash}m{1.3cm} >{\centering\arraybackslash}m{1.5cm} >   {\centering\arraybackslash}m{1.5cm} >{\centering\arraybackslash}m{1.5cm} @{}}
\toprule
\textbf{Datasets} & \textbf{Methods} & \textbf{Recall(\%)} & \textbf{Precision(\%)} & \textbf{F1(\%)} \\
\midrule
\multirow{6}{1.5cm}{\centering CoderEval} & \textcolor{tab-green}{DeepSeek-R1} & 8.75 & 6.25 & 7.29 \\
& Gemini-2.0 & 2.34 & 2.34 & 2.34 \\
& \textcolor{tab-blue}{o1-2024-12-17} & 0.61 & 0.61\% & 0.61 \\
& \textcolor{tab-green}{MAD\_DGO} & 42.50 (\textcolor{red}{$\uparrow$33.75}) & 12.50 (\textcolor{red}{$\uparrow$6.25})  & 19.32 (\textcolor{red}{$\uparrow$12.03})  \\
& MAD\_GDO & 29.24(\textcolor{red}{$\uparrow$26.90})  & 10.53 (\textcolor{red}{$\uparrow$8.19})  & 15.48 (\textcolor{red}{$\uparrow$13.14})  \\
& \textcolor{tab-blue}{MAD\_OGD} & 43.90(\textcolor{red}{$\uparrow$43.29})  & 5.49 (\textcolor{red}{$\uparrow$4.88})  & 9.76(\textcolor{red}{$\uparrow$9.15})  \\
\midrule
\multirow{6}{1.5cm}{\centering SWE-bench-NF} & \textcolor{tab-green}{DeepSeek-R1} & 23.64 & 20.91 & 22.19 \\
& Gemini-2.0 & 11.54 & 9.62 & 10.49 \\
& \textcolor{tab-blue}{o1-2024-12-17} & 5.83 & 5.83 & 5.83 \\
& \textcolor{tab-green}{MAD\_DGO} & 46.36(\textcolor{red}{$\uparrow$22.72})  & 20.18(\textcolor{red}{$\downarrow$0.73}) & 28.12(\textcolor{red}{$\uparrow$5.93}) \\
& MAD\_GDO & 38.46(\textcolor{red}{$\uparrow$26.92}) & 17.31(\textcolor{red}{$\uparrow$7.69}) & 23.87(\textcolor{red}{$\uparrow$13.38}) \\
& \textcolor{tab-blue}{MAD\_OGD} & 48.54(\textcolor{red}{$\uparrow$42.71}) & 4.85(\textcolor{red}{$\downarrow$0.98}) & 8.83(\textcolor{red}{$\uparrow$3.00}) \\
\bottomrule
\end{tabular}
\label{tab:adjusted_table}
\vspace{-0.3cm}
\end{table}


The MAD framework demonstrates superior performance in detecting low-quality CoTs, achieving significant advantages over any single LLM method in both Recall and the comprehensive F1-score. 
However, this performance advantage stems from the framework's inherent design of multi-agent interaction, which also gives rise to significant computational overhead, making the trade-off between performance and cost a key challenge for its practical application.




\begin{tcolorbox}[colback=gray!20, colframe=lightgray, boxrule=1pt, arc=5pt,left=5pt,right=5pt,top=5pt,bottom=5pt]
\textbf{Finding 4:} The MAD framework significantly outperforms single-model approaches for detecting low-quality CoTs, demonstrating exceptional performance, particularly in Recall and F1-score. However, the trade-off between the framework's superior performance and its inherent computational cost warrants further investigation for practical application.

\end{tcolorbox}
\section{LLMs' Self-Repair on Low-Quality CoTs}

\subsection{Study Design}

\textbf{Prompt Construction.}
Building on Section 4, where LLMs demonstrated the ability to detect low-quality CoTs, this section investigates their autonomous repair capability for identified low-quality CoTs.
Unlike the detection phase, we do not employ the MAD framework for the repair task, as it is inherently a single-LLM self-correction mechanism. 
Therefore, we focus on evaluating the self-repair capabilities of a single LLM under three granular feedback mechanisms---\textbf{Simple Feedback}, \textbf{Error Type Feedback}, and \textbf{Detailed Error Feedback}---to examine autonomous repair of CoT quality issues.
Error type feedback provides the LLM with information about the types of CoT quality issues, but does not offer detailed error specifics. For example, if the steps show inconsistency with the intended requirements or if the CoT is incomplete, we would provide error-type feedback.
Compared to the first two types of feedback, detailed error feedback provides specific information about the errors in the CoT. 
That is, during the repair process, we take the requirements, the original CoT, and the corresponding error information as input to let LLM repair the CoT.







%
\begin{tcolorbox}[
    colback=black!5,  
    colframe=black!75, 
    title=\textbf{Simple Feedback}, 
    fonttitle=\bfseries, 
    arc=2mm,
]
The generated steps have quality issues. Please fix the steps based on the requirement description.\par
\end{tcolorbox}

\begin{tcolorbox}[
    colback=black!5,  
    colframe=black!75, 
    title=\textbf{Error Type Feedback}, 
    fonttitle=\bfseries,
    arc=2mm,
]
The generated steps have quality issues. 
The CoT may have problems such as :

[ERROR TYPE 1] [BRIEFLY DEFINE ERROR TYPE 1],

[ERROR TYPE 2] [BRIEFLY DEFINE ERROR TYPE 2],

[ERROR TYPE 3] [BRIEFLY DEFINE ERROR TYPE 3],

\dots.

Original requirement: [REQUIREMENT DESCRIPTION]

Please fix the steps based on the requirement description.\par

\end{tcolorbox}

\begin{tcolorbox}[
    colback=black!5,  
    colframe=black!75, 
    title=\textbf{Detailed Error Feedback}, 
    fonttitle=\bfseries,
    arc=2mm,
]


The generated steps have quality issues. Please fix the steps according to the information below.

Original requirement: [REQUIREMENT DESCRIPTION]

Original CoT: [ORIGINAL CoT]

Detailed error information: [ERROR INFORMATION]

\end{tcolorbox}


Building on our earlier findings, where high-quality CoT reasoning is likely to yield correct code, we argue that successfully repaired CoTs can similarly produce accurate code. To evaluate the efficacy of CoT repair, we measure the correctness of the resulting code using the pass@k metric, a widely adopted benchmark for assessing code generation quality \cite{chen2022codet,liu2024your,yeo2024framework}. 
For each problem requiring repair, we generated one code solution.
If any of the solutions passed all test cases, the problem was considered resolved. This method systematically assesses the LLM's ability to refine and enhance the CoT using varied feedback types.

\subsection{Results}
\begin{table}[t!]
\caption{Repair results of different LLMs}
\label{Table: Repair results of different LLMs}
\centering
\scriptsize
\setlength{\tabcolsep}{3pt}
\renewcommand{\arraystretch}{1.2}
\begin{tabular}{@{} >{\centering\arraybackslash}m{2.3cm} >{\centering\arraybackslash}m{1.5cm} >{\centering\arraybackslash}m{1.2cm} >{\centering\arraybackslash}m{1.2cm} >{\centering\arraybackslash}m{1.5cm} @{}}
\toprule
\textbf{Feedback Type}\rule{0pt}{12pt} & \textbf{Datasets}\rule{0pt}{12pt} & \multicolumn{3}{c}{\textbf{Model (pass@1(\%))}} \\
\cmidrule(lr){3-5}
& & DeepSeek-R1 & Gemini-2.0 & o1-2024-12-17 \\
\midrule
\multirow{2}{2.5cm}{\centering Simple Feedback} & CoderEval & 1.7 & 5.2 & 3.0 \\
& SWE-bench-NF & 0.0 & 0.0 & 0.0 \\
\midrule
\multirow{2}{2.5cm}{\centering Error Type Feedback} & CoderEval & 2.2(\textcolor{red}{ $\uparrow$0.5}) & 5.7( \textcolor{red}{ $\uparrow$0.5 }) & 3.0( \textcolor{red} { -- }) \\
& SWE-bench-NF & 0.9(\textcolor{red}{ $\uparrow$0.9 }) & 0.9(\textcolor{red}{ $\uparrow$0.9 }) & 1.0(\textcolor{red}{ $\uparrow$1.0 }) \\
\midrule
\multirow{2}{2.5cm}{\centering Detailed Error Feedback} & CoderEval & 4.4(\textcolor{red}{ $\uparrow$2.2 }) & 7.4(\textcolor{red}{ $\uparrow$1.7 }) & 3.9(\textcolor{red}{ $\uparrow$0.9 }) \\
& SWE-bench-NF & 1.9(\textcolor{red}{ $\uparrow $1.0 }) & 2.8(\textcolor{red}{ $\uparrow$1.9 }) & 1.0( \textcolor{red} { -- }) \\
\bottomrule
\end{tabular}
\vspace{-0.35cm}
\end{table}

The section evaluates the ability of Gemini-2.0-Flash-Thinking-Exp-01-21, DeepSeek-R1, and o1-2024-12-17 to automatically repairing CoT quality issues arising from internal factors.
Before the repair, the code generated due to CoT internal quality issues failed to pass the test cases, resulting in pass@1 values all being 0\%. 
The experimental results after the repair are shown in Table~\ref{Table: Repair results of different LLMs}.
All three models demonstrate varying degrees of effectiveness on both CoderEval and SWE-bench-NF.
Gemini-2.0-Flash-Thinking demonstrates the highest performance, particularly exhibiting a significant advantage in the CoderEval benchmark.
Simple, error-type, and detailed error feedback achieve pass@1 of 5.2\%, 5.7\%, and 7.4\% respectively, significantly outperforming DeepSeek-R1 (1.7\%, 2.2\%, 4.4\%) and o1 (3.0\%, 3.0\%, 3.9\%).
In contrast, the SWE-bench-NF dataset posed a greater challenge for all models. 
Notably, simple feedback conditions substantially constrain models' ability to correct erroneous CoT.
Although progressively detailed feedback reveals an increasing trend in pass@1, the peak performance remains critically low at only 2.8\% (achieved by Gemini-2.0-Flash-Thinking under detailed error feedback).
This trend is particularly pronounced on CoderEval, while the improvement on SWE-bench-NF remained limited. 
We hypothesize this disparity arises because SWE-bench-NF derives from real GitHub issues and pull requests, where resolving these issues necessitates coordinated changes across multiple functions, classes, or files.
The provision of merely three feedback types proves inadequate for such multidimensional complexity.
Furthermore, DeepSeek-R1 exhibits a 2.7\% performance gain when transitioning from simple feedback to detailed error feedback on CoderEval—surpassing both Gemini-2.0-Flash-Thinking (2.2\%) and o1 (0.9\%). 
Conversely, on SWE-bench-NF, o1 demonstrates marginal superiority with a 1.0\% improvement under error-type feedback, outperforming other models (0.9\%).

In summary, these findings show that enhancing the design of feedback—such as incorporating detailed error feedback—can effectively boost model performance. 
However, precise contextual information localization is pivotal, particularly in repository-level code generation scenarios.
Additionally, the feedback mechanism employed by Gemini-2.0-Flash-Thinking in tackling complex code generation tasks merits further investigation.

\begin{tcolorbox}[colback=gray!20, colframe=lightgray, boxrule=1pt, arc=5pt,left=5pt,right=5pt,top=5pt,bottom=5pt]
\textbf{Finding 5:} 
Preliminary studies indicate that the self-repair capabilities of LLMs are limited, with the Gemini model exhibiting comparatively superior self-repair performance. Furthermore, during the repair process, the level of detail in feedback positively impacts the model’s repair performance.


\end{tcolorbox}





\section{Discussion}
\label{sec: Dis}
This section explores the implications of our findings for key stakeholders—developers, researchers, and LLM providers —and identifies the threats to the validity of our research.

\subsection{Implications of Findings}

\textbf{Developers:}
Our findings serve as a critical reminder that integrating LLMs into the development workflow is not a passive act of delegation, but rather a collaborative, human-in-the-loop process that demands scrutiny and critical validation.
The prevalence of missing contextual information and unclear requirements (Finding 1) underscores that the responsibility for quality control begins with the prompt itself \cite{10.1145/3660810}. 
Crafting effective prompts is a pivotal step in optimizing human-AI collaboration.
Developers should not only provide explicit instructions but also meticulously articulate the implicit context that a human collaborator would naturally infer.

Furthermore, our analysis reveals a treacherous disconnect between the model's reasoning steps and its final code.
The fact that 11.9\% of CoTs exhibited quality issues despite the corresponding code being correct, while 18.5\% of correct CoTs still resulted in erroneous code (Finding 3), is particularly telling.
This insight serves as a stark warning that the correctness of a model's reasoning process does not guarantee the correctness of its final output. 
Therefore, developers should adopt a ``dual verification" approach, independently scrutinizing both the CoTs and the code's functionality. 
Relying on the model's self-repair capabilities is also a limited strategy.
While detailed feedback can yield better results (Finding 5), making human oversight by the developer a critical final step for validation.

\textbf{Researchers:} 
Our study on LLM-driven code generation reveals critical gaps in model capabilities that point toward promising research directions for the code intelligence community.
First, our findings—that a majority of generation failures stem from a lack of precise context and an inability to grasp implicit requirements (Findings 1 \& 2)—directly highlight a core limitation in current models: a lack of accurate context localization and deep inferential capabilities. Therefore, developing advanced Retrieval-Augmented Generation (RAG) techniques tailored for code repositories and enhancing models' ability to perform sophisticated reasoning on the retrieved context are directions worth exploring \cite{xia2025demystifying,li2024codetree,chen2025locagent}.
Second, our findings reveal the fundamental inconsistency between a model's CoT and its final code output (Finding 3). \textit{What, then, is the true nature of a CoT? Should it be interpreted as an authentic representation of the model's internal reasoning, or as a convincing narrative generated independently of the process that created the code?}  
This fundamental question opens a critical line of inquiry, prompting researchers to investigate the true causal mechanisms of LLM-based program synthesis and develop new methods to enforce alignment between a model's articulated reasoning and its functional output.
Finally, our study indicates that the self-repair capabilities of LLMs are limited, though repair performance is positively impacted by the level of detail in user feedback (Finding 5). This points to a frontier in interactive program repair \cite{tang2024code,liu2024fastfixer}. 
As such, formalizing the interactive feedback loop and designing models that can transition from brittle auto-correction to robust, conversational debugging is an interesting direction that is worth exploring.

\textbf{LLM Providers:} 
For LLM providers, a key strategic challenge is not merely the correctness of the final code, but the trustworthiness of the generation process itself \cite{turpin2023language,arcuschin2025chain}. Our findings reveal a critical disconnect between the model's stated reasoning and its output. This implies that simply generating a plausible CoT is insufficient; providers must invest in ensuring procedural fidelity—that the model actually follows its articulated reasoning steps. This will help companies enhance product competitiveness and drive sustained market growth.

\subsection{Threats to Validity}
\textbf{Threats to external validity} pertain to the generalizability of our findings and the benchmark used.
Our empirical study, along with the selected benchmark, specifically focuses on Python code generation tasks.
Python is a widely studied language within the code generation community and is commonly used in benchmarking practices.
This prompted us to conduct a thorough investigation into the quality issues surrounding LLM-generated CoT for Python code generation.
Furthermore, it would be valuable to extend this analysis to other programming languages to evaluate the quality of LLM-generated CoT in a broader context.
Given the general applicability of our analysis procedure, it could be readily adapted and applied to various programming languages beyond Python, providing a more comprehensive understanding of CoT quality across different contexts.

\textbf{Threats to internal validity} might be introduced from our manual analysis and taxonomy construction.
Evaluating the quality of CoT is somewhat subjective, and different labelers may have varying determinations of the same CoT samples.
To address this issue, we created a preliminary categorization system for each category, and two of the authors were responsible for the verification process.
Additionally, we organized discussions and meetings to resolve any conflicts or uncertainties that arose. 
This comprehensive process significantly improved the reliability and quality of our manual analysis of the factors affecting the taxonomy of CoT quality in LLM-generated code.

\textbf{Threats to construct validity} relate to the suitability of our evaluation of the experiments on identifying and mitigating factors affecting CoT quality. The evaluation results may be sensitive to the prompt format. To mitigate this issue, we conducted small-scale preliminary testing, experimenting with different prompts to select the prompts that yield the most consistent and optimal performance across various models.
\setlength{\parskip}{0pt}






\section{Conclusion}
\label{sec:conclusion}
In this study, we conduct a comprehensive empirical analysis of the factors affecting the quality of CoTs generated by LLMs and develop a taxonomy that systematically categorizes these factors into external factors (e.g., unclear implementation details) and internal factors (e.g., misunderstanding explicit requirements). 
These findings highlight the complex challenges involved in ensuring high-quality CoTs for reliable code generation.
Additionally, we explore the subtle relationship between CoT quality and code correctness.
Notably, even when a CoT is entirely accurate, it does not necessarily generate correct code, revealing the gap between reasoning and implementation. 
Conversely, even when a CoT is flawed, it sometimes generates correct code, suggesting that LLMs may rely on their inherent knowledge to compensate for reasoning errors.
This duality reveals both the limitations and potential of LLMs in this domain.
Building on this, we preliminarily explore the potential of LLMs in self-detecting and self-repairing erroneous CoTs.
Experimental results demonstrate that the MAD framework's error detection significantly outperforms single-LLM approaches. 
Furthermore, integrating detailed error descriptions into CoT repair procedures substantially improves LLMs' error correction capabilities.


\section{Data Availability}
We provide the data and scripts online to facilitate replication or future work at: \url{https://github.com/binquanzzz/CoT-Eval}



\bibliography{ref}



\end{document}